\documentclass[prx,english,notitlepage,twocolumn,superscriptaddress]{revtex4-2}
\usepackage{amssymb}
\usepackage{hyperref}
\usepackage{graphicx}
\usepackage{amsmath}
\usepackage{color}
\usepackage{float}
\usepackage{xcolor}
\pagecolor{white}

\renewcommand{\arraystretch}{2.0}

\makeatletter
\renewcommand*\env@matrix[1][\arraystretch]{%
  \edef\arraystretch{#1}%
  \hskip -\arraycolsep
  \let\@ifnextchar\new@ifnextchar
  \array{*\c@MaxMatrixCols c}}
\makeatother

\newcommand{\COMMENTED}[1]{}

\begin{document}

\title{Quantum Weyl-Heisenberg antiferromagnet }
\author{Peter Rosenberg}
\affiliation{Département de Physique \& Institut Quantique, Université de Sherbrooke, Québec, Canada J1K 2R1}
\author{Efstratios Manousakis}
\affiliation{Department of Physics, Florida State University, Tallahassee, Florida 32306, USA}
\affiliation{Department of Physics, National and Kapodistrian University of Athens, Panepistimioupolis, Zografos, 157 84 Athens, Greece}

\begin{abstract} 
 Beginning from the conventional square-lattice nearest-neighbor antiferromagnetic Heisenberg
 model, we allow the $J_x$ and $J_y$ couplings to be anisotropic, with their values depending on the
 bond orientation. The emergence of anisotropic, bond-dependent, couplings should be expected
 to occur naturally in most antiferromagnetic compounds which undergo structural
 transitions that reduce the point-group symmetry at lower temperature.
 Using the spin-wave approximation, we study the model in several parameter
 regimes by diagonalizing the reduced Hamiltonian exactly, and computing the edge spectrum and Berry connection vector, 
 which show clear evidence of localized topological charges.
We discover phases that exhibit Weyl-type spin-wave dispersion, characterized by pairs of degenerate points 
and edge states, as well as phases supporting lines of degeneracy. We also identify a parameter regime in which
there is an exotic state hosting gapless linear spin-wave dispersions with different longitudinal and transverse spin-wave velocities. 
\end{abstract}

\maketitle

\section{Introduction}
More than three decades ago, Anderson proposed~\cite{Anderson} that the
large-amplitude quantum-spin fluctuations, present in the ground-state of the lowest-possible spin, the spin-1/2, Heisenberg antiferromagnet on the low-dimensional lattice, the square lattice, destroy the N\'eel order, and that the new emerging state of matter
could be the foundation of the superconductivity observed in the cuprates.  Subsequent
work showed that the idea was not applicable to the spin-1/2 square lattice Heisenberg
antiferromagnet~\cite{Manousakis-RMP}. However, Anderson's proposal for the resonating-valence-bond state, whose original inception dates back
nearly half a century~\cite{ANDERSON1973153}, served as
the ignition of a new field of research in ``quantum magnetism'', aimed at
realizing the quantum-spin liquid state. Most ensuing attempts to realize
such states have focused on low-spin low-dimensionality systems, but they have departed from
bi-partite lattices in an effort to introduce geometric frustration or frustrating interactions.

In the meantime, topology has emerged as a means of characterizing electronic
structure~\cite{RevModPhys.82.3045}, introducing new concepts and descriptions for condensed matter systems,
such as the topological insulator~\cite{doi:10.1126/science.1133734,PhysRevLett.98.106803}, the topological superconductor~\cite{Bernevig} and the Weyl semi-metal~\cite{RevModPhys.90.015001,PhysRevB.83.205101}.
Importing these ideas from topology into the field of quantum magnetism
is a compelling pursuit for various reasons, including technological
applications. For example, quantum
spins may be used as the degrees of freedom in spintronics-magnonics applications~\cite{PhysRevLett.127.247202} and in many
proposals for quantum computing devices that require long coherence time-scales. Such long life-times are expected to be a fundamental property of topological quantum-spin states because of their inherently robust nature.

There are various ways to construct quantum-spin models
furnishing topological characteristics.
In this work, we focus on a model that
exhibits quantum-spin Weyl behavior, an idea that has been explored in the literature in the context
of various models.
Quantum-spin Weyl behavior has been found in a pyrochlore antiferromagnet where a local spin anisotropic coupling~\cite{Li2016,PhysRevB.97.115162}, allowed by the symmetry group, led to the emergence of Weyl magnons.
Weyl magnons have also been discussed in the case of Honeycomb lattice ferromagnets~\cite{PhysRevB.104.104419} such as in CsI$_3$ ~\cite{PhysRevX.8.041028}.
Tunable topological magnon phases in
triangular-honeycomb lattices have been studied~\cite{PhysRevB.107.024408}
and also predicted~\cite{PhysRevB.107.174404} in layered ferrimagnets, where they arise as a result of spin-orbit coupling. Rather recently,
an anomalous thermal Hall effect in the insulating van der Waals magnet
VI$_3$ was observed~\cite{PhysRevLett.127.247202}, its existence attributed to topological magnons.
Furthermore, it has been shown~\cite{annurev}  that the addition of a second nearest-neighbor Dzyaloshinskii–Moriya
interaction to the isotropic Heisenberg interaction can create Weyl magnons. 
Topological magnetic excitations have also been
discussed~\cite{PhysRevB.98.060405} in the context of the Kitaev-Heisenberg model.
Lastly, the general topological features of such bosonic models have also been discussed in the literature
~\cite{PhysRevB.102.125127}.

In the present paper, we propose that such exotic, difficult-to-realize, models are not necessary
to observe quantum-spin Weyl behavior. In fact, as we will show, this behavior is present in a slight 
(and known) modification of the familiar, garden-variety,
spin-1/2 Heisenberg XYZ antiferromagnet on the square lattice!
Beginning from the conventional square-lattice XYZ antiferromagnetic Heisenberg
model, we introduce anisotropic couplings $J_x$ and $J_y$ whose values depend on the
bond orientation. Within the spin-wave approximation, the effective bosonic model
exhibits a Weyl-type spin-wave dispersion with edge states. The model is first investigated
analytically, followed by a numerical investigation~\cite{PhysRevB.106.054511,PhysRevB.104.134511}
of the edge spectrum and Berry connection vector in
various parameter regimes, which show clear evidence of localized topological charges.
In short, we demonstrate that breaking the inversion symmetry introduces
various interesting topological excitations of Weyl-boson character with
topologically non-trivial edge states carrying topological charge.
 
Finally, we argue that such Weyl-magnon
behavior, in perhaps the most basic model of magnetism, i.e., the nearest-neighbor Heisenberg 
antiferromagnet, with no additional terms, should be ubiquitous. We discuss how to 
experimentally realize such square-lattice quantum magnets endowed with the topological excitations 
and edge-states found here.

The paper is organized as follows. In the following section we introduce and analyze
our model. In Sec.~\ref{results} we present our results and in Sec.~\ref{conclusion} we discuss 
our conclusions and how to realize the proposed behavior experimentally.

\section{Model}
\label{sec:latt_ham}

The Hamiltonian for the familiar $XYZ$ quantum Heisenberg antiferromagnet can be written as:
\begin{align}
  \hat H = \sum_{\alpha,\langle ij \rangle}J_\alpha(\boldsymbol{\delta}_{ij}) \hat S^\alpha_i \hat S^\alpha_j - h \sum_{i=1}^N \hat{S}^z_i,
\end{align}
where $\alpha=x,y,z$, $\boldsymbol{\delta}_{ij}$ is the
vector connecting the nearest-neighbors $ij$, and $N$ is the number of sites. We will assume $J_z \geq J_x, J_y$, and that there is $(\pi,\pi)$ antiferromagnetic order along the $z$ internal-spin direction. We define,
  \begin{eqnarray}
    J_{\pm}(\boldsymbol{\delta}_{ij}) &\equiv& {{J_x(\boldsymbol{\delta}_{ij}) \pm J_y(\boldsymbol{\delta}_{ij})} \over 2},
  \end{eqnarray}
  and introduce the spin-ladder operators,
   \begin{eqnarray}
   \hat{S}^x_i = {{\hat{S}^+_i+\hat{S}^-_j} \over 2}, \hskip 0.2 in 
   \hat{S}^y_i = {{\hat{S}^+_i-\hat{S}^-_j} \over {2 i}},
 \end{eqnarray}
  with which the Hamiltonian can be rewritten in the following form:
  
 \begin{eqnarray}
   \hat H &=&  \sum_{{\bf R} \in A ,\boldsymbol \delta} \left\{J_z \hat S^z_{\bf R} \hat S^z_{\bf R+ \boldsymbol \delta}  + {{1} \over 2}
J_+(\boldsymbol \delta) (\hat S^+_{\bf R}
  \hat S^-_{\bf R + \boldsymbol \delta} + \hat S^-_{\bf R} \hat S^+_{\bf R + \boldsymbol \delta} )\right.\notag\\
  &+& \left.{{ 1 } \over 2} J_-(\boldsymbol \delta) (\hat S^+_{\bf R}
  \hat S^+_{\bf R + \boldsymbol \delta} + \hat S^-_{\bf R}   \hat S^-_{\bf R + \boldsymbol \delta})\right\}
  - h \sum_{i=1}^N \hat S^z_i,
\end{eqnarray}
 where we have introduced two sublattices, $A$ and $B$. The sum over $\bf R$ only includes sites on the $A$ sublattice, and the vector ${\boldsymbol{\delta}}= \pm a \hat x,\pm a \hat y$ connects sites on the $A$ sublattice to sites on the $B$ sublattice. The sums over $\bf R$ and ${\boldsymbol{\delta}}$ therefore include all sites of the lattice.
  
\begin{figure}[!ht]
\begin{center}
\includegraphics[width=0.65\columnwidth]{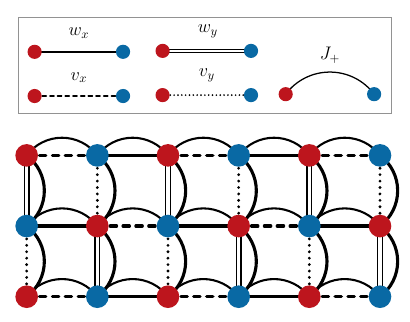}
\caption{Illustration of lattice with coupling terms. Sublattice $A$ is denoted by the red circles,
and sublattice $B$ by the blue circles. The box above the illustration indicates the line style
corresponding to each coupling.}
\label{fig:latt}
\end{center}
\end{figure}

Next, we carry out the well-known~\cite{Manousakis-RMP} substitution of the spin operators in terms of boson operators,
$\hat{a}_{\bf{R}}$ ($\hat{a}_{\bf{R}}^\dagger$) and $\hat{b}_{\bf R}$ ($\hat{b}_{\bf R}^\dagger$) that destroy (create) spin fluctuations
on the $A$ and $B$ sublattice respectively.
Then, by expanding the Hamiltonian and keeping terms up
to quadratic order in these boson operators we obtain:
 \begin{eqnarray}
   \hat H_{LSW} &=& -{2 N } S^2 J_z +
   \varepsilon_A \sum_{{\bf R}\in A}^N \hat{a}^{\dagger}_{\bf R}
   \hat{a}_{\bf R}
   + \varepsilon_B\sum_{{\bf R}\in B}^N \hat{b}^{\dagger}_{\bf R} \hat{b}_{\bf R} \nonumber \\
    &+& {J_+} S \sum_{{\bf R} \in A, \boldsymbol \delta} 
  (\hat{a}_{\bf R} \hat{b}_{{\bf R} + \boldsymbol \delta} + \hat{a}^{\dagger}_{\bf R} \hat{b}^{\dagger}_{{\bf R} + \boldsymbol \delta})\notag\\
 &+& S \sum_{{\bf R}\in A,\alpha}\left[v_\alpha 
    (\hat{a}^{\dagger}_{\bf R} \hat{b}_{{\bf R} + \hat \alpha a} + \hat{a}_{\bf R} \hat{b}^{\dagger}_{{\bf R} + \hat \alpha a})\right. \notag\\
    &&\quad\quad+ w_\alpha \left.
  (\hat{a}^{\dagger}_{\bf R} \hat{b}_{{\bf R} - \hat \alpha a} + \hat{a}_{\bf R} \hat{b}^{\dagger}_{{\bf R} - \hat \alpha a})\right],
  \end{eqnarray}
  where $\alpha = x,y$, the spin $S=1/2$, and we have taken $J_z$ and $J_+$ to be directionally-independent, i.e. $J_z(\boldsymbol \delta)=J_z$, $J_+(\boldsymbol \delta)=J_+$, and let 
$J_-(a\hat x) = v_x$, $J_-(-a\hat x) = w_x$, $J_-(a\hat y) = v_y$, and $J_-(-a\hat y) = w_y$. In addition we have defined, $\varepsilon_A \equiv 4 S J_z + h$, and $\varepsilon_B \equiv 4 S J_z- h$.
An illustration of the terms of the Hamiltonian is shown in Fig. ~\ref{fig:latt}.

In momentum space we have,
 \begin{align}
   \hat H_{\textrm{LSW}} &= -{2 N} S^2 J_z +
   \varepsilon_A \sum_{\mathbf{k}}  {\tilde a}^{\dagger}_{\mathbf{k}} {\tilde a}_{\mathbf{k}}
     + \varepsilon_B \sum_{\mathbf{k}}  {\tilde b}^{\dagger}_{\mathbf{k}} {\tilde b}_{\mathbf{k}}
     \nonumber \\
    &+ 4 S {J_+} \sum_{\mathbf{k}} \gamma(\mathbf{k})
     ({\tilde a}_{\mathbf{k}} {\tilde b}_{-\mathbf{k}} + {\tilde a}^{\dagger}_{\mathbf{k}}
     {\tilde b}^{\dagger}_{-\mathbf{k}})] \nonumber\\
  &+  S \sum_{\alpha, \mathbf{k}} (v_\alpha
    e^{i k_\alpha a} + w_\alpha e^{-i k_\alpha a}) {\tilde a}^{\dagger}_{\mathbf{k}} {\tilde b}_{\mathbf{k}} + h.c.,
 \end{align}
 where $\gamma(\mathbf{k}) \equiv {1 \over 2} (\cos(k_x a) + \cos(k_y a))$.
 In matrix form we have,
\begin{equation}
\hat{H}_{\textrm{LSW}}=   \frac{1}{2}\sum_{\mathbf{k}}
\begin{pmatrix}
\tilde{a}^{\dagger}_{\mathbf k} & \tilde{b}^{\dagger}_{\mathbf k} & \tilde{a}_{\mathbf{ -k}} & \tilde{b}_{\mathbf{-k}}   
\end{pmatrix}
\mathcal{H}(\mathbf{k})
\begin{pmatrix}
\tilde{a}_{\mathbf k} \\
\tilde{b}_{\mathbf k} \\
\tilde{a}^{\dagger}_{\mathbf{-k}} \\
\tilde{b}^{\dagger}_{\mathbf{-k}} 
\end{pmatrix},
\label{eq:LSW}
\end{equation}
where,
\begin{equation}
\mathcal{H}(\mathbf{k})=
\begin{pmatrix}
\varepsilon_A && V(\mathbf k) && 0 && \Delta(\mathbf k) \\
V^*(\mathbf k)  && \varepsilon_B && \Delta(\mathbf k) && 0\\
0 && \Delta(\mathbf k) && \varepsilon_A && V(\mathbf k) \\
\Delta(\mathbf k) && 0 && V^*(\mathbf k) && \varepsilon_B 
\end{pmatrix},
\label{eq:Hk}
\end{equation}
with $\Delta(\mathbf k) \equiv 4 S J_+ \gamma(\bf k)$, and
 $V(\mathbf k) \equiv  S [v_x e^{i k_x a} + w_x e^{-i k_x a} + v_y e^{i k_y a} + w_y e^{-i k_y a}]$.

As a simple illustration of the Weyl character of the model we consider the case
of $\Delta(\mathbf{k})=0$ and $h = 0$ (i.e. $\varepsilon_A=\varepsilon_B\equiv\varepsilon$). 
In this limit, the Hamiltonian in Eq.~(\ref{eq:Hk}) is composed of two identical $2\times2$ blocks. 
This $2\times2$ matrix can be written in the form $\varepsilon\mathbb{I}+\boldsymbol \tau \cdot {\bf h}({\bf k})$,
where  $\boldsymbol \tau$ is a vector of the Pauli matrices in the pseudo-spin basis and,
\begin{equation}
{\bf h}({\bf k}) \equiv
\begin{pmatrix}
V_R(\mathbf{k}) \\
V_I(\mathbf{k}) \\
0
\end{pmatrix},
\end{equation}
where $V_R(\mathbf{k})$ and $V_I(\mathbf{k})$ are the real an imaginary parts of 
$V(\mathbf{k})$, respectively.
The term $\varepsilon\mathbb{I}$ is a constant that shifts the overall energy of the system,
but does not affect its topology, therefore we omit it from the following. 
If we take as a particular example the case of $v_x=w_x$ and $v_y=-w_y$,
and expand near the nodal 
points, retaining terms below second order in $\mathbf{k}$ we obtain,
\begin{equation}
\boldsymbol\tau \cdot {\bf h}({\bf k}) \approx \alpha_x \Delta k_x \tau_x + \alpha_y \Delta k_y \tau_y,
\label{eq:H_Weyl}
\end{equation}
where $\alpha_x=-2v_xa$ and $\alpha_y= 2 v_ya$ are constants, and
$\Delta k_x \equiv (k_x-k^0_x)$, $\Delta k_y \equiv (k_y-k^0_y)$, where $(k^0_x,k^0_y)$ is the
momentum of the nodal point. Eq.~(\ref{eq:H_Weyl}) has the explicit form of the Weyl Hamiltonian.

In order to diagonalize this Hamiltonian we seek a transformation, $\mathbb{T}(\mathbf{k})$,  
commonly referred to as a paraunitary transformation \cite{COLPA1978327} that preserves the bosonic commutation relations:
\begin{equation}
\begin{pmatrix}
\tilde{a}^\dagger_\mathbf{k} & \tilde{b}^\dagger_\mathbf{k} & \tilde{a}_\mathbf{-k} & \tilde{b}_\mathbf{-k}
\end{pmatrix}
=
\begin{pmatrix}
\alpha^\dagger_\mathbf{k} & \beta^\dagger_\mathbf{k} & \alpha_\mathbf{-k} & \beta_\mathbf{-k}
\end{pmatrix}
\mathbb{T}_\mathbf{k}.
\end{equation}
This constraint implies the following for the Hamiltonian matrix,
\begin{equation}
\mathbb{T}^\dagger(\mathbf{k})
\mathcal{H}(\mathbf{k})
\mathbb{T}(\mathbf{k})=\mathbb{E}(\mathbf{k}),
\end{equation}
where $\mathbb{E}(\mathbf{k})$ is a diagonal matrix of the energy eigenvalues and,
\begin{equation}
\mathbb{T}^\dagger(\mathbf{k})
\tau_z
\mathbb{T}(\mathbf{k})=\mathbb{T}(\mathbf{k})
\tau_z
\mathbb{T}^\dagger(\mathbf{k})=\tau_z,
\end{equation}
where $\tau_z=\text{diag}\{1,1,-1,-1\}$.
The matrix, 
\begin{align}
\tilde{\mathcal{H}}(\mathbf{k})&\equiv\mathbb{T}^\dagger(\mathbf{k})
\mathcal{H}(\mathbf{k})
\mathbb{T}(\mathbf{k}),\notag\\
&=
\begin{pmatrix}
\varepsilon_A && V(\mathbf k) && 0 && \Delta(\mathbf k) \\
V^*(\mathbf k)  && \varepsilon_B && \Delta(\mathbf k) && 0\\
0 && -\Delta(\mathbf k) && -\varepsilon_A && -V(\mathbf k) \\
-\Delta(\mathbf k) && 0 && -V^*(\mathbf k) && -\varepsilon_B 
\end{pmatrix},
\end{align} can be diagonalized by a unitary transformation, which
yields the spin-wave dispersions.
In the case of periodic boundary conditions the
eigenenergies can be obtained analytically and expressed as:
\begin{align}
  \omega^{\pm}(\mathbf{k}) &= \sqrt{{{\varepsilon^2_A+\varepsilon^2_B} \over 2} + \vert V(\mathbf{k})\vert^2 - \vert\Delta(\mathbf{k})\vert^2 \pm \sqrt{Z(\mathbf{k})}} \\
Z(\mathbf{k}) &\equiv    \left[{{(\varepsilon_A-\varepsilon_B)^2} \over 2} + 2 \vert V(\mathbf{k})\vert^2\right] \left[{{(\varepsilon_A+\varepsilon_B)^2} \over 2} - 2 |\Delta(\mathbf{k})|^2\right] \notag\\
&\quad+ 4 \vert\Delta(\mathbf{k})\vert^2 V^2_R(\mathbf{k}).
  \end{align}

\section{Results}
\label{results}

As alluded to above, the presence of anisotropic couplings in this model can
lead to band structures with topological character, including Weyl points and lines of degeneracy.
The condition for these band touching points is given by the following:
\begin{equation}
V_R^2(\mathbf{k})\varepsilon^2+V_I^2(\mathbf{k})(\varepsilon^2-\Delta^2)=0.
\label{eq:nodal_cond}
\end{equation}

In the following we take $h=0$, which means $\varepsilon_A=\varepsilon_B\equiv\varepsilon$.
We note that the system is gapped when $h\neq 0$, but we focus here on the parameter regime that 
supports Weyl physics. Because we have chosen the $z$-axis to be along the direction of the antiferromagnetic order,
we assume that $J_z\geq J_+$. This implies that the term $(\varepsilon^2-\Delta^2)$ is always positive, 
which means that Eq.~(\ref{eq:nodal_cond}) can only be satisfied when $V(\mathbf{k})=0$. In terms of the
real and imaginary parts of $V(\mathbf{k})$ we have:
\begin{align}
(v_x+w_x)\cos(k_xa) +  (v_y+w_y)\cos(k_ya) &= 0, \label{eq:vr0} \\
(v_x-w_x)\sin(k_xa) +  (v_y-w_y)\sin(k_ya) &= 0. \label{eq:vi0}
\end{align}

The above system of equations can be solved by isolating $\cos(k_xa)$ (or $\cos(k_ya)$) in Eq.~(\ref{eq:vr0})
and substituting into Eq.~(\ref{eq:vi0}), or alternatively by isolating $\sin(k_xa)$ (or $\sin(k_ya)$) in Eq.~(\ref{eq:vi0})
and substituting into Eq.~(\ref{eq:vr0}). This general solution involves a division by
$v_x+w_x$ (or $v_y+w_y$), or $v_x-w_x$ (or $v_y-w_y$), respectively.
This division is not well-defined when any of these coefficients is equal to zero,
therefore these cases must be treated separately.
We give the solutions for these cases (1-6) in Table~\ref{table:1}.

Cases 1, 2, 4, and 5 correspond to solutions with Weyl points, while the remaining
cases correspond to solutions where the spin-wave dispersions are degenerate
along a line in momentum-space. In the remainder of this section we will illustrate
examples of these cases. Note that all parameters and energies are given in units
of $J_z$.

\begin{table}[!htb]

  \begin{tabular}{ |c|c|c|}
    \hline
Case & Parameters & Solutions \\
    \hline
 1&   $v_y=w_y$,  $v_x\ne w_x$ & $k_x =0,{{\pi}\over a},k_ya=\cos^{-1}\left(\mp{{v_x+w_x} \over {2v_y}}\right) $ \\
    \hline
    2&    $v_x=w_x$,  $v_y\ne w_y$ &
    $k_xa =\cos^{-1}\left(\mp {{v_y+w_y} \over {2 v_x}}\right),k_y=0,{{\pi}\over a} $ \\
    \hline
3 &    $v_x=w_x$,  $v_y= w_y$ & $\cos(k_xa) = -\frac{v_y}{v_x}\cos(k_ya)$ \\
    \hline
4 &    $v_y=-w_y$,  $v_x \ne -w_x$ & $k_x=\pm {{\pi}\over {2a}},k_ya=\sin^{-1}\left(-\frac{v_x-w_x}{2v_y}\right)$  \\
    \hline
5 &    $v_x=-w_x$,  $v_y \ne -w_y$ &$k_y=\pm {{\pi}\over {2a}},k_xa=\sin^{-1}\left(-\frac{v_y-w_y}{2v_x}\right)$  \\
    \hline
6 &    $v_x=-w_x$,  $v_y=-w_y$ & $\sin(k_xa) = -\frac{v_y}{v_x}\sin(k_ya)$\\
\hline    
7 &    $\alpha=1$,  $\beta=1$ & $k_ya = \pm \pi + k_xa$\\
\hline    
8 &    $\alpha=1$,  $\beta=-1$ & $k_ya = \pm \pi -k_xa$\\
\hline    
9 &    $\alpha=-1$,  $\beta=1$ & $k_y = -k_x$\\
\hline    
10 &    $\alpha=-1$,  $\beta=-1$ & $k_y = k_x$\\
\hline    
  \end{tabular}
  \caption{Solutions to Eqs.~(\ref{eq:vr0}),~(\ref{eq:vi0}) for various parameter choices.}
		\label{table:1}

\end{table}

\begin{figure*}[!ht]
\begin{center}
\includegraphics[width=0.9\textwidth]{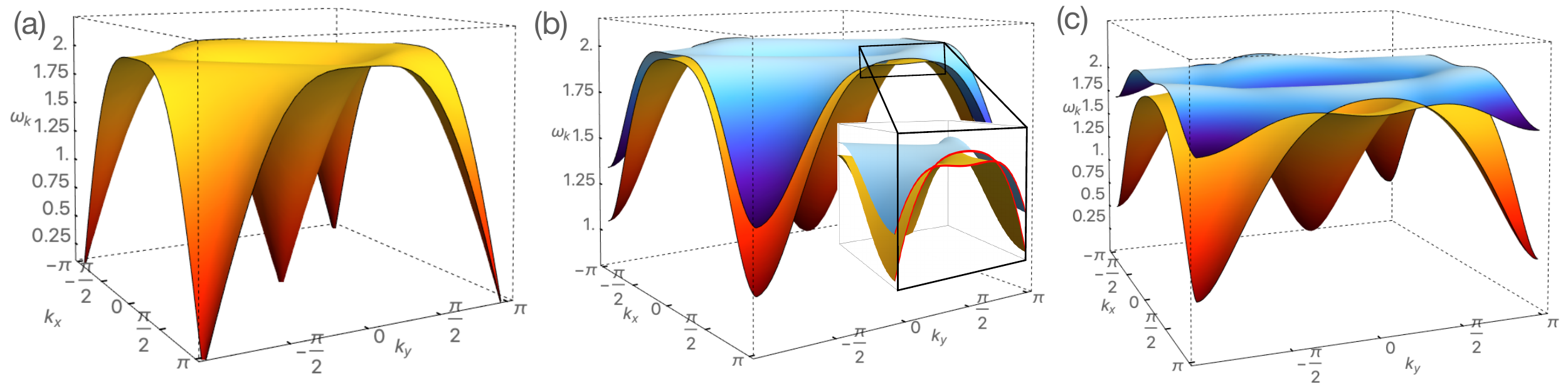}
\caption{Spin-wave dispersion for (a) $v_x=w_x=v_y=w_y=0.0$, $J_+=1.0$, (b)
$v_x=0.05$, $w_x=0.025$, $v_y=0.05$, $w_y=0.05$, $J_+=0.8$, 
(c) $v_x=0.2$, $w_x=0.05$, $v_y=0.2$, $w_y=0.2$, $J_+=0.8$. The inset of (b) highlights
a pair of degenerate points that emerge with the introduction of anisotropic couplings. Another pair
of degenerate points along the $k_x$-axis is not visible in the figure.}
\label{fig:iso_to_small_aniso}
\end{center}
\end{figure*}

\begin{figure*}[!ht]
\begin{center}
\includegraphics[width=0.9\textwidth]{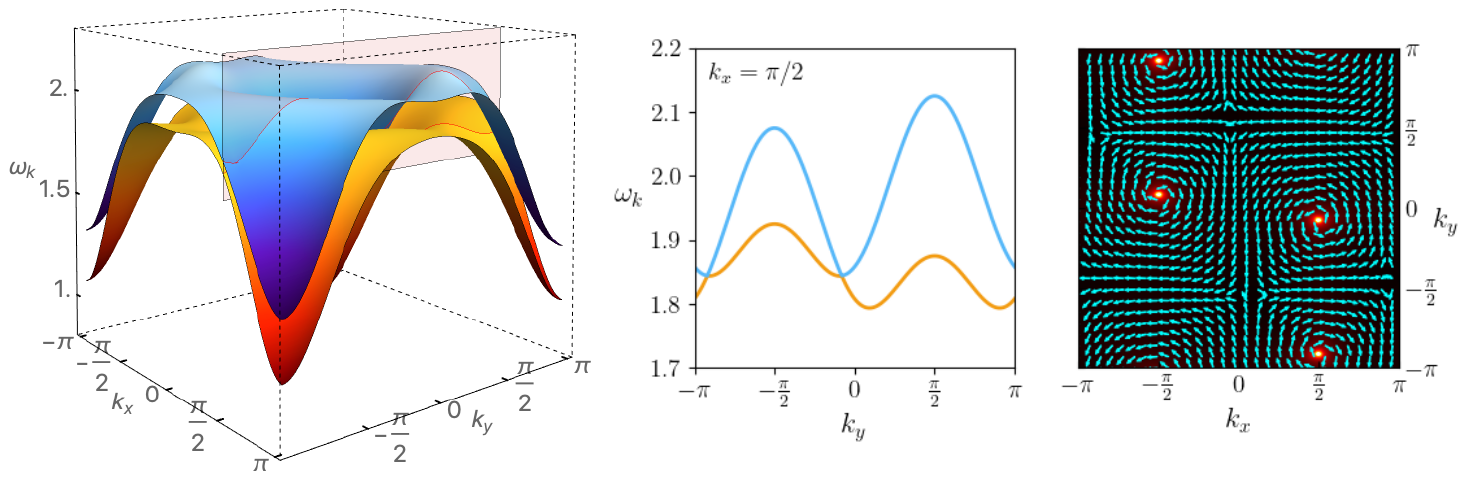}
\caption{(left) Spin-wave dispersion for $v_x=0.1$, $w_x=0.05$, $v_y=0.1, w_y=-0.1$, $J_+=0.8$. (center) Cut along $k_x=\pi/2$ (indicated by red region in left panel). (right) Berry connection vector. The direction is indicated by the arrows while the color plot shows the magnitude.} 
 \label{fig:case4}
\end{center}
\end{figure*}

\begin{figure*}[!ht]
\begin{center}
\includegraphics[width=0.9\textwidth]{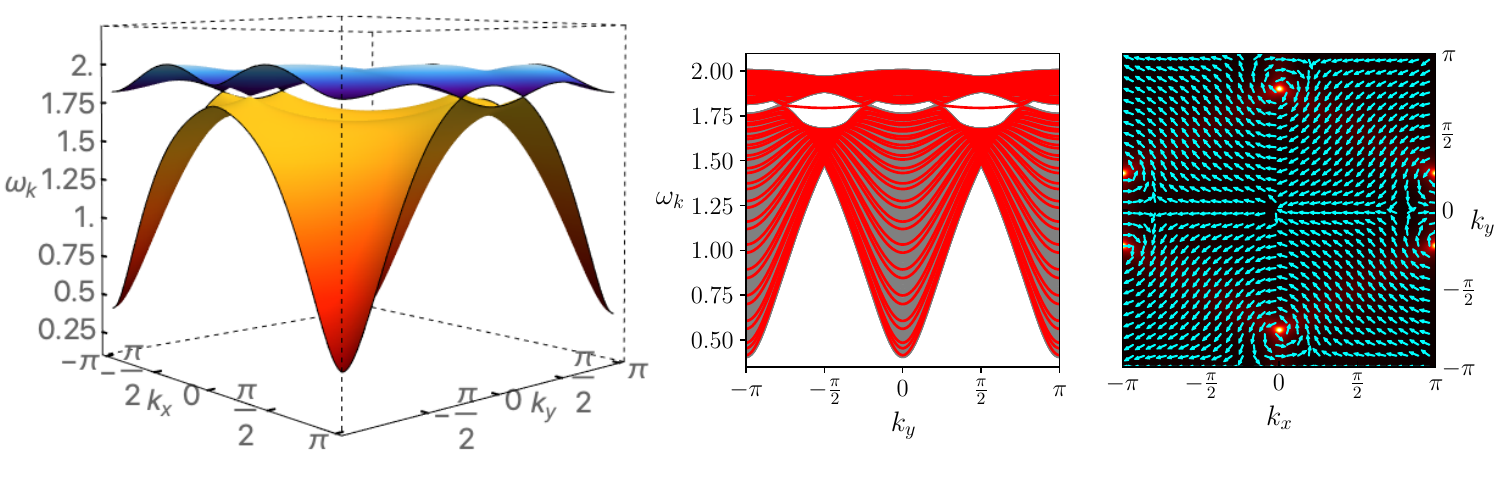}
\caption{(left) Spin-wave dispersion for $v_x=0.364$, $w_x=0.0455$, $v_y=0.273, w_y=0.273$, $J_+=0.723$. (center) Edge spectrum, bulk shown in gray. (right) Berry connection vector. The direction is indicated by the arrows while the color plot shows the magnitude.} 
 \label{fig:large-anisotropy1}
\end{center}
\end{figure*}

\begin{figure*}[!ht]
\begin{center}
\includegraphics[width=0.9\textwidth]{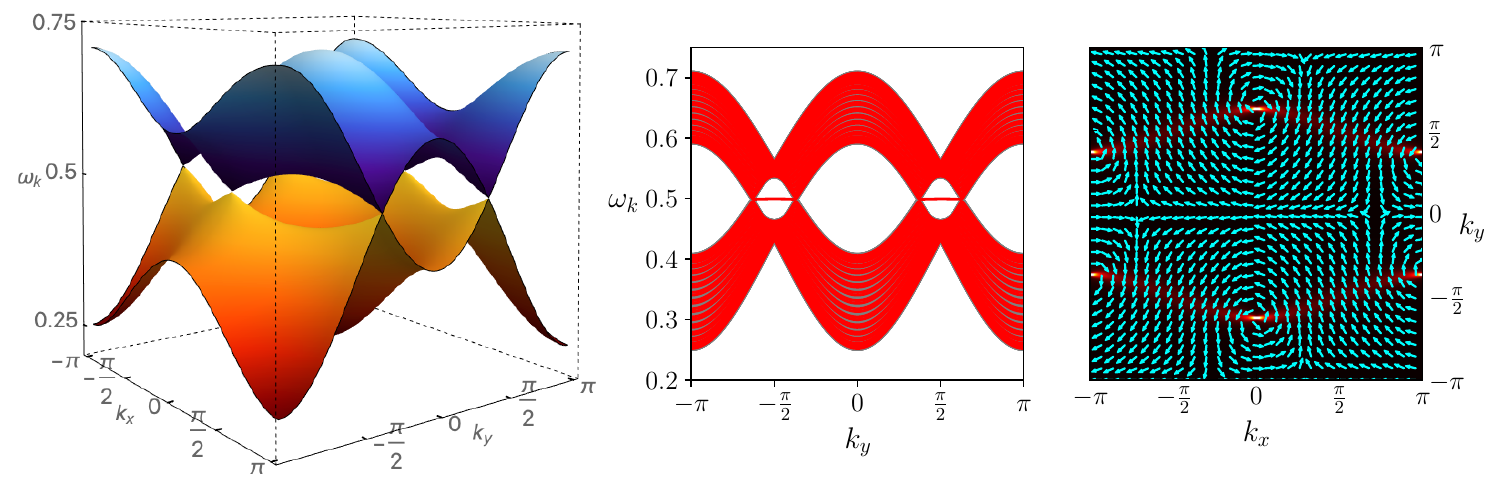}
\caption{(left) Spin-wave dispersion for $v_x=0.4$, $w_x=0.1$, $v_y=w_y=0.625$, $J_+=0.25$. (center) Edge spectrum, bulk shown in gray. (right) Berry connection vector. The direction is indicated by the arrows while the color plot shows the magnitude.} 
 \label{fig:large-anisotropy2}
\end{center}
\end{figure*}

\begin{figure*}[!ht]
\begin{center}
\includegraphics[width=0.9\textwidth]{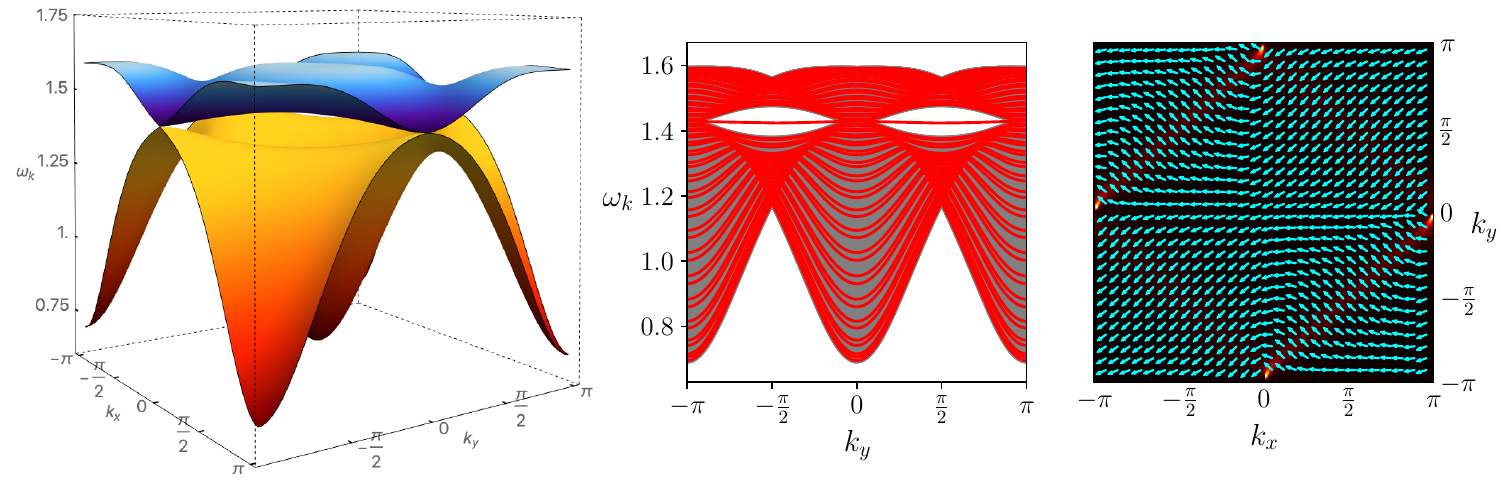}
\caption{(left) Spin-wave dispersion for $v_x=0.357$, $w_x=0.143$, $v_y=0.286$, $w_y=0.214$, $J_+=0.571$.  (center) Edge spectrum, bulk shown in gray. (right) Berry connection vector. The direction is indicated by the arrows while the color plot shows the magnitude.}
 \label{fig:special-case}
\end{center}
\end{figure*}

\begin{figure*}[!ht]
\begin{center}
\includegraphics[width=0.9\textwidth]{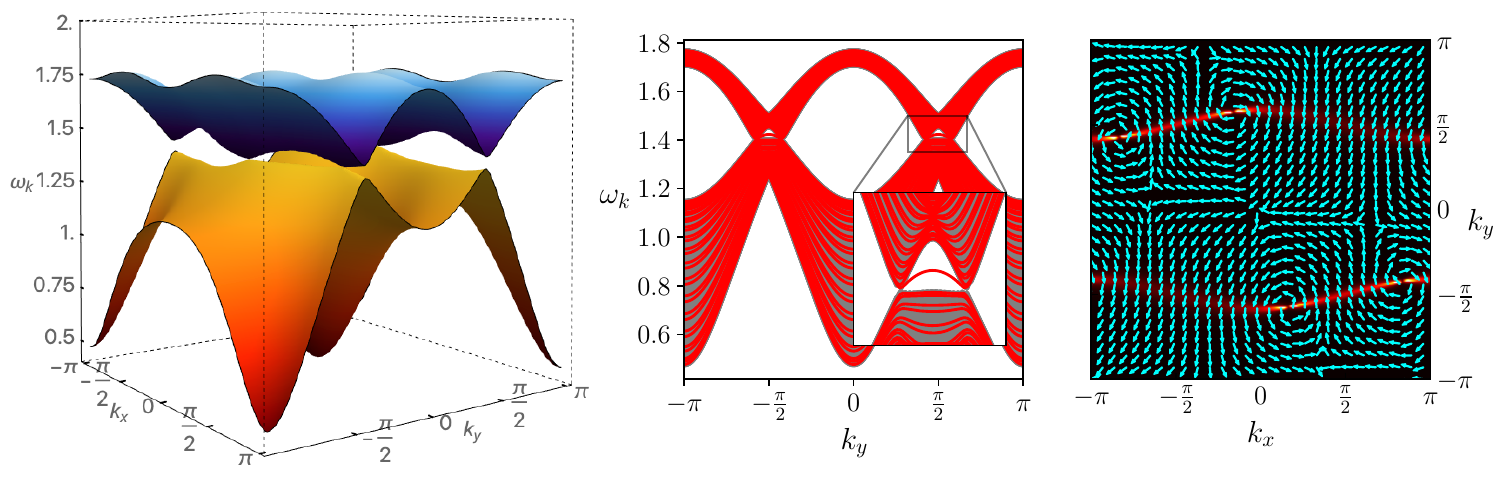}
\caption{(left) Spin-wave dispersion for $v_x=0.214$, $w_x=0.071$, $v_y=0.571$, $w_y=0.5$, $J_+=0.571$.  (center) Edge spectrum, bulk shown in gray. (right) Berry connection vector. The direction is indicated by the arrows while the color plot shows the magnitude.}
\label{fig:dispersive-edge-state}
\end{center}
\end{figure*}

In addition to the cases highlighted above (1-6 in Table~\ref{table:1}), we note
four additional special cases. First, if none of the cases 1-6 occur, then the
coefficients $v_x\pm w_x \neq 0$ and $v_y\pm w_y \neq 0$, so we can rewrite 
Eqs.~(\ref{eq:vr0}) as,
\begin{align}
\cos(k_xa) =& -\alpha\cos(k_ya), \label{eq:alpha} \\
\sin(k_xa)   =& -\beta \sin(k_ya), \label{eq:beta} \\
\end{align}
where we have defined,
\begin{align}
 \beta \equiv&  \,{{v_y-w_y} \over {v_x-w_x}},\\
\alpha \equiv& \,{{v_y+w_y}\over {v_x+w_x}}.
\end{align}
The solutions corresponding to the four combinations of $\alpha =\pm 1$, $\beta = \pm 1$,
which represent lines of degeneracy, are given as cases 7-10 in Table~\ref{table:1}.

When none of these ten special cases occurs, i.e, when $\vert v_y \vert \neq \vert w_y \vert$ and $\vert v_x \vert \neq \vert w_x \vert$,
and $\alpha \ne 1$ (and $\beta \ne 1$),
the general solution to the system of Eqs.~(\ref{eq:vr0}),~(\ref{eq:vi0}) gives the momenta at which Weyl-points occur:
\begin{align}
k_xa &= \cos^{-1}[\mp \alpha \Lambda]\label{eq:general-solution1} \\
k_ya &= \cos^{-1}[\pm\Lambda], \label{eq:general-solution2}
\end{align}
where $\Lambda \equiv \sqrt{(\beta^2-1)/(\beta^2-\alpha^2)}$.

Having classified the various phases exhibiting topological features,
including Weyl points and lines of degeneracy, we proceed by 
presenting examples of the band structure, edge spectrum, and
Berry connection (defined as $\mathbf{A} = -i\langle \psi_0 \vert \boldsymbol\nabla_\mathbf{k}\vert \psi_0 \rangle$, where $\vert \psi_0\rangle$
is the ground state), for these different states.

\subsection{Solutions with Weyl-points}

We begin by considering cases 1, 2, 4, and 5 of Table~\ref{table:1}, as well
as the general case given by Eqs.~(\ref{eq:general-solution1}),~(\ref{eq:general-solution2}), 
all of which yield spin-wave spectra with Weyl points.

We first present the case of isotropic coupling ($J_+=J_z=1.0$, with all other parameters equal to zero), which corresponds to the well-known nearest-neighbor antiferromagnetic Heisenberg model on the square-lattice, in order to provide a comparison for the spin-wave dispersions for the cases with anisotropic couplings. Fig.~\ref{fig:iso_to_small_aniso}(a) shows the spin-wave dispersion for the isotropic case, which displays the well-known linear momentum dependence near $\mathbf{k}=0$ and at $\mathbf{k}  = (\pm \pi/a,\pm \pi/a)$. In this case, the bands associated with $\omega^+$ and $\omega^-$ are degenerate. The introduction of coupling terms between the two sublattices ($v_\alpha, w_\alpha$) breaks this degeneracy. If $J_+<J_z$ a gap opens
at $\Gamma$ (and $(\pm \pi/a,\pm \pi/a)$), which can lead to non-trivial topological band structures.

Figs.~\ref{fig:iso_to_small_aniso}(b)-(c)
illustrate the presence of pairs of Weyl-point solutions, given by
case 1 of Table~\ref{table:1}, for small values of $v_x, v_y, w_x, w_y$, i.e., for small
deviations from the isotropic limit.
Note that case 2 is equivalent to case 1 with the $x$ and $y$ axes interchanged.
Both of these cases represent situations where the inversion symmetry is broken in 
either the $x$ or $y$ direction.
 
Fig.~\ref{fig:case4} shows an example of case 4. Notice that
there are two pairs of Weyl points at the location given in
Table~\ref{table:1}, which are seen as singularities of the Berry-connection
vector shown in Fig~\ref{fig:case4}(c). Case 5 is obtained from case 4 by
interchanging the $x$ and $y$ axes.

We continue by exploring several different parameter sets,
with larger deviations from the isotropic limit, that display unique topological features. 
Fig.~\ref{fig:large-anisotropy1} and Fig.~\ref{fig:large-anisotropy2}
give the calculated spin-wave dispersion (left),
edge spectrum superimposed on the bulk states, which are projected as
a gray background ribbon (center), and the Berry connection vector (right).
To compute the edge spectrum we treat a system with semi-open boundary
conditions that has $N_x = 50$ layers along the $x$-direction and is periodic
along the $y$-direction.
In both cases, there are edge states that connect the bulk Weyl points
in the lower (upper) half of the Brillouin zone at $k_x = 0$ to those in the lower (upper) half of
the Brillouin zone at $k_x = \pi$. The right panel demonstrates that
there is a $ 2\pi$ counter-clockwise and a $2\pi$ clockwise rotation of
the Berry connection vector as we travel around the positive and negative pseudo-helicity Weyl point,
respectively.

Fig.~\ref{fig:special-case} again shows the calculated spin-wave dispersion, the
edge spectrum superimposed on the bulk states, and the Berry connection 
vector (using the same notations as in Fig.~\ref{fig:large-anisotropy1})
for a special case of parameters which gives $\alpha=1$ but $\beta \ne \pm 1$,
meaning degenerate points exist at $k_x=0$, $k_y a = \pm \pi$ and vice versa.
In this case, we find a quadratic dispersion at the gapless band touching point. 
In this parameter regime, when $\alpha > 1$ there are two pairs of Weyl points that
merge at $k_x=0$, $k_y a = \pm \pi$ and $k_xa= \pm \pi$, $k_y = 0$ as $\alpha \rightarrow 1$, 
yielding the quadratic band touching point we observe. For $\alpha < 1$ there is no solution to 
Eqs.~(\ref{eq:alpha}),(\ref{eq:beta}), so the system is gapped.

Fig.~\ref{fig:dispersive-edge-state} illustrates a situation where the anisotropy is large and 
as a result the edge-state has significant dispersion along the $\mathbf{k}$
direction parallel to the edge (in this case $k_y$). This case is similar to that shown in Fig.~\ref{fig:large-anisotropy2}, 
however here $v_y \neq w_y$, which causes the Weyl points to shift away from $k_x = \pm\pi$ and $k_x = 0$, and
leads to a dispersive edge-state.

\subsection{Solutions with lines of degeneracy}

In this subsection we consider parameter sets that lead to solutions with lines 
of degeneracy; these correspond to cases 3, and 6-10 of Table~\ref{table:1}. 

\begin{figure}[!ht]
\begin{center}
\includegraphics[width=\columnwidth]{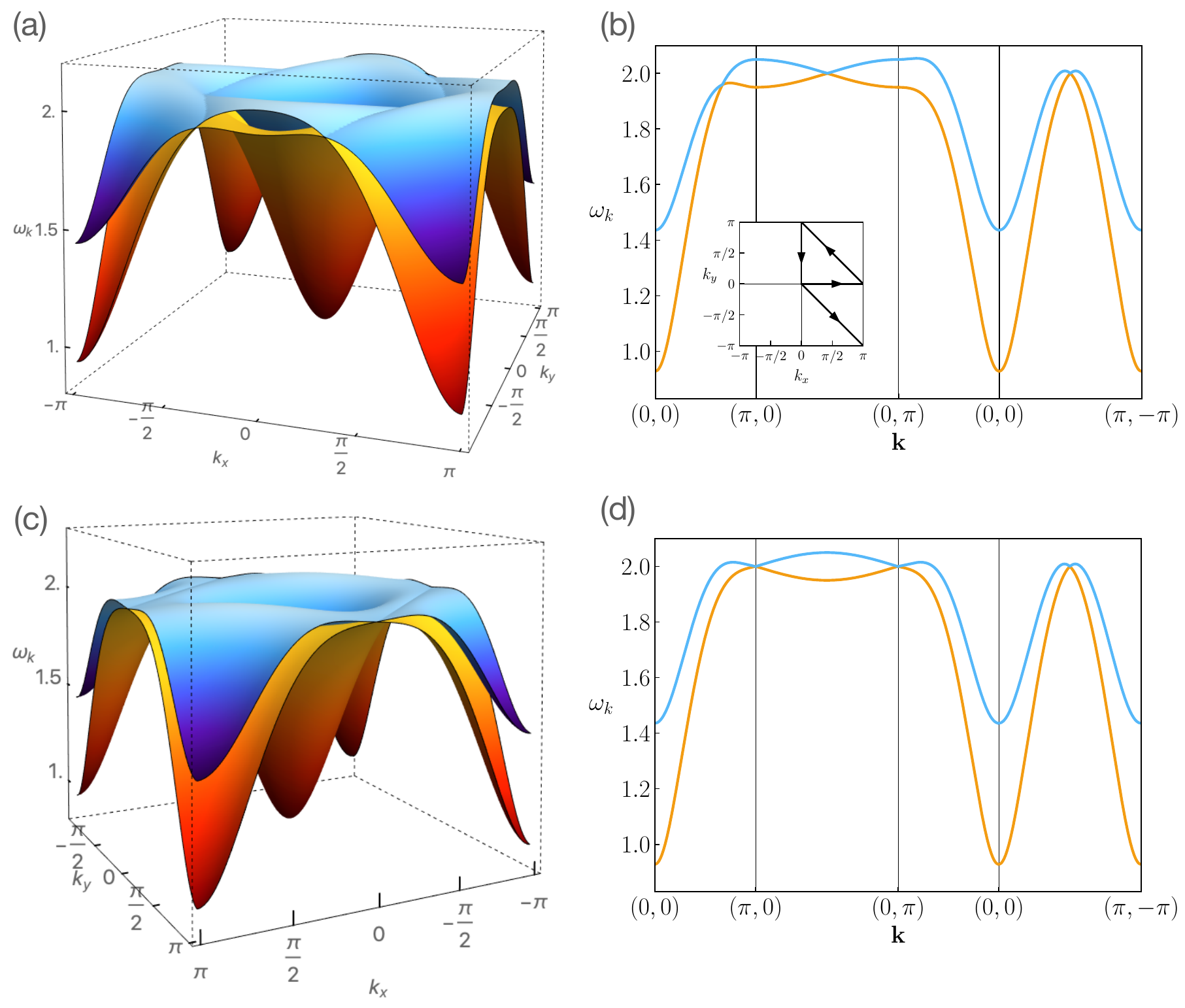}
\caption{(a) Spin-wave dispersion for $v_x=0.1$, $w_x=0.1$, $v_y=0.05$, $w_y=0.05$, $J_+=0.8$.
(b) Path through Brillouin zone (as shown in inset) for parameters in (a).
(c) Spin-wave dispersion for $v_x=0.1$, $w_x=0.05$, $v_y=0.1$, $w_y=0.05$, $J_+=0.8$.
(d) Path through Brillouin zone for parameters in (c).}
\label{fig:line_degeneracies1}
\end{center}
\end{figure}

Fig.~\ref{fig:line_degeneracies1}(a) illustrates case 3 of
Table~\ref{table:1}, while Fig.~\ref{fig:line_degeneracies1}(c) presents
an example of case 7, where $\alpha =1$ and $\beta=1$, in which case, there is a pair of lines of degeneracy
at $k_y = \pm \pi + k_x$. Figs.~\ref{fig:line_degeneracies1}(b),(d) show
the spin-wave dispersions along the path in the Brillouin zone indicated in the
inset of Fig.~\ref{fig:line_degeneracies1}(b), which crosses the lines of degeneracy
in several locations.

\begin{figure}[!ht]
\begin{center}
\includegraphics[width=\columnwidth]{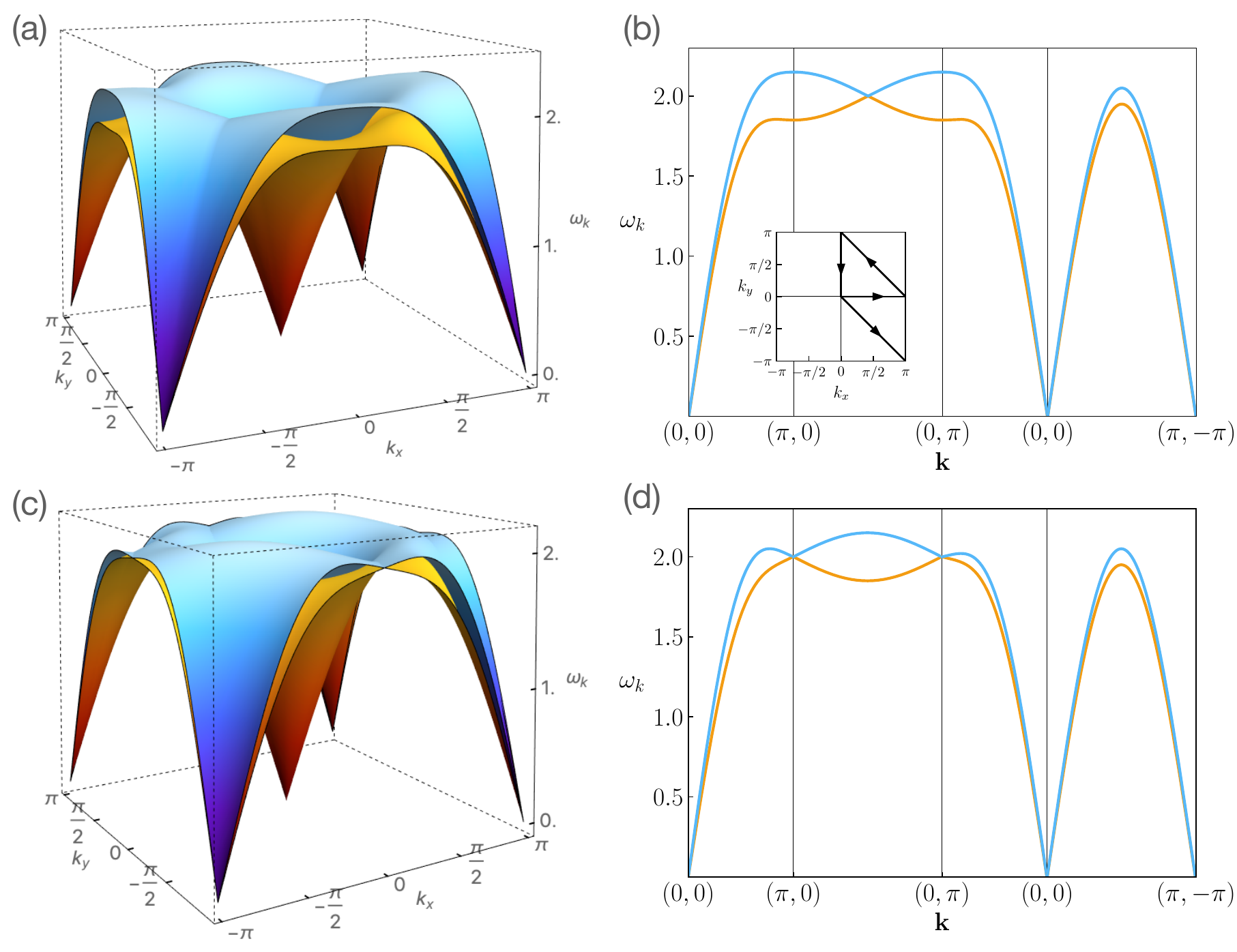}
\caption{(a) Spin-wave dispersion for $v_x=0.1$, $w_x=0.05$, $v_y=-0.1$, $w_y=-0.05$, $J_+=1.0$.
(b) Path through Brillouin zone (as shown in inset) for parameters in (a).
(c) Spin-wave dispersion for $v_x=0.1$, $w_x=-0.1$, $v_y=0.05$, $w_y=-0.05$, $J_+=1.0$.
(d) Path through Brillouin zone for parameters in (c)}
\label{fig:line_degeneracies2}
\end{center}
\end{figure}

Finally, we consider cases 10 and 6 of Table~\ref{table:1}, which also correspond to
lines of degeneracy.  Fig.~\ref{fig:line_degeneracies2}(a) is an example
of case 10 where $\alpha=\beta=-1$, which results in a line of degeneracy
at $k_y=k_x$. Fig.~\ref{fig:line_degeneracies2}(c) is an example
of case 6 where the couplings $v_x$ and $v_y$ in the $+x$ and $+y$ directions respectively are 
opposite to those in the $-x$ and $-y$ directions. Note that in both of
these cases the global degeneracy present in the isotropic limit is broken. Moreover, the spin-wave 
dispersions are linear near $\mathbf{k}=0$ and $\mathbf{k}  = (\pm \pi/a,\pm \pi/a)$, suggesting a pair of 
spin waves with different velocities along the longitudinal and transverse directions. This behavior is evident 
in Figs.~\ref{fig:line_degeneracies2}(b),(d), which show the spin-wave dispersions along the path in the Brillouin 
zone indicated in the inset of Fig.~\ref{fig:line_degeneracies2}(b).
 
\section{Conclusion}
\label{conclusion}

In the present work, we have considered a slight modification of the 
familiar isotropic Heisenberg antiferromagnet on
a square lattice, a model that is believed to describe a variety of
non-geometrically frustrated spin systems, either as is, or as
the foundation of Hamiltonians that describe a vast set of magnetic systems.

We have discovered that by allowing the $J_x$ and $J_y$ coupling constants
to differ from $J_z$, so that their values become slightly different for different orientations of
the nearest-neighbor bonds, thereby breaking the inversion symmetry of the lattice, the model 
can host interesting magnetic excitations with fascinating topological
character.

We are able to analytically solve the modified model, within the
spin-wave approximation, i.e., by taking into account the role
of quantum-spin fluctuations around a N\'eel ordered state.
The magnetic Bravais-lattice unit-cell remains the same as in
the simpler case of the antiferromagnetic order, however, because our
model breaks inversion symmetry, the degeneracy of the magnon
dispersion is lifted and we obtain a pair of magnon dispersions which
can cross at lines of degeneracy, or form singular pairs of Weyl points
with opposite topological charge associated with a pseudo-helicity.
In order to characterize the topological nature of these singularities,
we have calculated the edge spectrum and the field of the
Berry connection vector.
We find that the model can host pairs of Weyl-magnon states with
opposite topological charge connected by
Weyl-arc edge states. In addition, we find parameter regimes
 characterized by lines of degeneracy.

Given that the nearest-neighbor Heisenberg antiferromagnet
finds ubiquitous application in various magnetic materials, our findings
may be easily realizable in nature.
In fact, the slight modifications between the $x$-$y$ and $z$ components
of the Hamiltonian discussed in this work are expected to be necessary 
to describe the magnetic excitations of a wide variety of
two-dimensional antiferromagnets. In general, such anisotropic couplings  
should be expected when the antiferromagnet undergoes structural
transition at lower temperature to phases of lower symmetry,
such as orthorhombic or monoclinic structures.
In a similar fashion, Peierls distortion can lead to different values for the $J_-$ coupling
in the forward and backward bonds along a given direction, due to 
differences in bond length caused by the distortion.
The parent compounds of the cuprate superconductors, for example La$_2$CuO$_4$,
undergo a tetragonal to orthorhombic transition at lower temperature. In the orthorhombic phase,
the structural distortions of the 2D Cu-O lattice should modify the
couplings in such a way that this or other parent compounds
of the family of oxides could host some of the states found in the present work.

In addition, continual progress in the field of ultra-cold atoms
suggests that our model could be artificially engineered in optical
lattices, where the conventional version of such a model has already been
simulated \cite{Simon2011,doi:10.1126/science.aam7838}.

\section*{Acknowledgments}
E. M. acknowledges support by the U.S. National Science Foundation under Grant No. NSF-EPM-2110814


\begin{thebibliography}{24}%
\makeatletter
\providecommand \@ifxundefined [1]{%
 \@ifx{#1\undefined}
}%
\providecommand \@ifnum [1]{%
 \ifnum #1\expandafter \@firstoftwo
 \else \expandafter \@secondoftwo
 \fi
}%
\providecommand \@ifx [1]{%
 \ifx #1\expandafter \@firstoftwo
 \else \expandafter \@secondoftwo
 \fi
}%
\providecommand \natexlab [1]{#1}%
\providecommand \enquote  [1]{``#1''}%
\providecommand \bibnamefont  [1]{#1}%
\providecommand \bibfnamefont [1]{#1}%
\providecommand \citenamefont [1]{#1}%
\providecommand \href@noop [0]{\@secondoftwo}%
\providecommand \href [0]{\begingroup \@sanitize@url \@href}%
\providecommand \@href[1]{\@@startlink{#1}\@@href}%
\providecommand \@@href[1]{\endgroup#1\@@endlink}%
\providecommand \@sanitize@url [0]{\catcode `\\12\catcode `\$12\catcode
  `\&12\catcode `\#12\catcode `\^12\catcode `\_12\catcode `\%12\relax}%
\providecommand \@@startlink[1]{}%
\providecommand \@@endlink[0]{}%
\providecommand \url  [0]{\begingroup\@sanitize@url \@url }%
\providecommand \@url [1]{\endgroup\@href {#1}{\urlprefix }}%
\providecommand \urlprefix  [0]{URL }%
\providecommand \Eprint [0]{\href }%
\providecommand \doibase [0]{https://doi.org/}%
\providecommand \selectlanguage [0]{\@gobble}%
\providecommand \bibinfo  [0]{\@secondoftwo}%
\providecommand \bibfield  [0]{\@secondoftwo}%
\providecommand \translation [1]{[#1]}%
\providecommand \BibitemOpen [0]{}%
\providecommand \bibitemStop [0]{}%
\providecommand \bibitemNoStop [0]{.\EOS\space}%
\providecommand \EOS [0]{\spacefactor3000\relax}%
\providecommand \BibitemShut  [1]{\csname bibitem#1\endcsname}%
\let\auto@bib@innerbib\@empty
\bibitem [{\citenamefont {Anderson}(1987)}]{Anderson}%
  \BibitemOpen
  \bibfield  {author} {\bibinfo {author} {\bibfnamefont {P.~W.}\ \bibnamefont
  {Anderson}},\ }\bibfield  {title} {\bibinfo {title} {{The Resonating Valence
  Bond State in La$_2$CuO$_4$ and Superconductivity}},\ }\href
  {https://doi.org/10.1126/science.235.4793.1196} {\bibfield  {journal}
  {\bibinfo  {journal} {Science}\ }\textbf {\bibinfo {volume} {235}},\ \bibinfo
  {pages} {1196} (\bibinfo {year} {1987})}\BibitemShut {NoStop}%
\bibitem [{\citenamefont {Manousakis}(1991)}]{Manousakis-RMP}%
  \BibitemOpen
  \bibfield  {author} {\bibinfo {author} {\bibfnamefont {E.}~\bibnamefont
  {Manousakis}},\ }\bibfield  {title} {\bibinfo {title} {{The
  spin-\textonehalf{} Heisenberg antiferromagnet on a square lattice and its
  application to the cuprous oxides}},\ }\href
  {https://doi.org/10.1103/RevModPhys.63.1} {\bibfield  {journal} {\bibinfo
  {journal} {Rev. Mod. Phys.}\ }\textbf {\bibinfo {volume} {63}},\ \bibinfo
  {pages} {1} (\bibinfo {year} {1991})}\BibitemShut {NoStop}%
\bibitem [{\citenamefont {Anderson}(1973)}]{ANDERSON1973153}%
  \BibitemOpen
  \bibfield  {author} {\bibinfo {author} {\bibfnamefont {P.}~\bibnamefont
  {Anderson}},\ }\bibfield  {title} {\bibinfo {title} {{Resonating valence
  bonds: A new kind of insulator?}},\ }\href
  {https://doi.org/https://doi.org/10.1016/0025-5408(73)90167-0} {\bibfield
  {journal} {\bibinfo  {journal} {Materials Research Bulletin}\ }\textbf
  {\bibinfo {volume} {8}},\ \bibinfo {pages} {153} (\bibinfo {year}
  {1973})}\BibitemShut {NoStop}%
\bibitem [{\citenamefont {Hasan}\ and\ \citenamefont
  {Kane}(2010)}]{RevModPhys.82.3045}%
  \BibitemOpen
  \bibfield  {author} {\bibinfo {author} {\bibfnamefont {M.~Z.}\ \bibnamefont
  {Hasan}}\ and\ \bibinfo {author} {\bibfnamefont {C.~L.}\ \bibnamefont
  {Kane}},\ }\bibfield  {title} {\bibinfo {title} {{Colloquium: Topological
  insulators}},\ }\href {https://doi.org/10.1103/RevModPhys.82.3045} {\bibfield
   {journal} {\bibinfo  {journal} {Rev. Mod. Phys.}\ }\textbf {\bibinfo
  {volume} {82}},\ \bibinfo {pages} {3045} (\bibinfo {year}
  {2010})}\BibitemShut {NoStop}%
\bibitem [{\citenamefont {Bernevig}\ \emph {et~al.}(2006)\citenamefont
  {Bernevig}, \citenamefont {Hughes},\ and\ \citenamefont
  {Zhang}}]{doi:10.1126/science.1133734}%
  \BibitemOpen
  \bibfield  {author} {\bibinfo {author} {\bibfnamefont {B.~A.}\ \bibnamefont
  {Bernevig}}, \bibinfo {author} {\bibfnamefont {T.~L.}\ \bibnamefont
  {Hughes}},\ and\ \bibinfo {author} {\bibfnamefont {S.-C.}\ \bibnamefont
  {Zhang}},\ }\bibfield  {title} {\bibinfo {title} {{Quantum Spin Hall Effect
  and Topological Phase Transition in HgTe Quantum Wells}},\ }\href
  {https://doi.org/10.1126/science.1133734} {\bibfield  {journal} {\bibinfo
  {journal} {Science}\ }\textbf {\bibinfo {volume} {314}},\ \bibinfo {pages}
  {1757} (\bibinfo {year} {2006})}\BibitemShut {NoStop}%
\bibitem [{\citenamefont {Fu}\ \emph {et~al.}(2007)\citenamefont {Fu},
  \citenamefont {Kane},\ and\ \citenamefont {Mele}}]{PhysRevLett.98.106803}%
  \BibitemOpen
  \bibfield  {author} {\bibinfo {author} {\bibfnamefont {L.}~\bibnamefont
  {Fu}}, \bibinfo {author} {\bibfnamefont {C.~L.}\ \bibnamefont {Kane}},\ and\
  \bibinfo {author} {\bibfnamefont {E.~J.}\ \bibnamefont {Mele}},\ }\bibfield
  {title} {\bibinfo {title} {{Topological Insulators in Three Dimensions}},\
  }\href {https://doi.org/10.1103/PhysRevLett.98.106803} {\bibfield  {journal}
  {\bibinfo  {journal} {Phys. Rev. Lett.}\ }\textbf {\bibinfo {volume} {98}},\
  \bibinfo {pages} {106803} (\bibinfo {year} {2007})}\BibitemShut {NoStop}%
\bibitem [{\citenamefont {Bernevig}\ and\ \citenamefont
  {Hughes}(2013)}]{Bernevig}%
  \BibitemOpen
  \bibfield  {author} {\bibinfo {author} {\bibfnamefont {B.~A.}\ \bibnamefont
  {Bernevig}}\ and\ \bibinfo {author} {\bibfnamefont {T.~L.}\ \bibnamefont
  {Hughes}},\ }\href@noop {} {\emph {\bibinfo {title} {Topological insulators
  and topological superconductors}}}\ (\bibinfo  {publisher} {Princeton
  University Press},\ \bibinfo {address} {Princeton},\ \bibinfo {year}
  {2013})\BibitemShut {NoStop}%
\bibitem [{\citenamefont {Armitage}\ \emph {et~al.}(2018)\citenamefont
  {Armitage}, \citenamefont {Mele},\ and\ \citenamefont
  {Vishwanath}}]{RevModPhys.90.015001}%
  \BibitemOpen
  \bibfield  {author} {\bibinfo {author} {\bibfnamefont {N.~P.}\ \bibnamefont
  {Armitage}}, \bibinfo {author} {\bibfnamefont {E.~J.}\ \bibnamefont {Mele}},\
  and\ \bibinfo {author} {\bibfnamefont {A.}~\bibnamefont {Vishwanath}},\
  }\bibfield  {title} {\bibinfo {title} {{Weyl and Dirac semimetals in
  three-dimensional solids}},\ }\href
  {https://doi.org/10.1103/RevModPhys.90.015001} {\bibfield  {journal}
  {\bibinfo  {journal} {Rev. Mod. Phys.}\ }\textbf {\bibinfo {volume} {90}},\
  \bibinfo {pages} {015001} (\bibinfo {year} {2018})}\BibitemShut {NoStop}%
\bibitem [{\citenamefont {Wan}\ \emph {et~al.}(2011)\citenamefont {Wan},
  \citenamefont {Turner}, \citenamefont {Vishwanath},\ and\ \citenamefont
  {Savrasov}}]{PhysRevB.83.205101}%
  \BibitemOpen
  \bibfield  {author} {\bibinfo {author} {\bibfnamefont {X.}~\bibnamefont
  {Wan}}, \bibinfo {author} {\bibfnamefont {A.~M.}\ \bibnamefont {Turner}},
  \bibinfo {author} {\bibfnamefont {A.}~\bibnamefont {Vishwanath}},\ and\
  \bibinfo {author} {\bibfnamefont {S.~Y.}\ \bibnamefont {Savrasov}},\
  }\bibfield  {title} {\bibinfo {title} {{Topological semimetal and Fermi-arc
  surface states in the electronic structure of pyrochlore iridates}},\ }\href
  {https://doi.org/10.1103/PhysRevB.83.205101} {\bibfield  {journal} {\bibinfo
  {journal} {Phys. Rev. B}\ }\textbf {\bibinfo {volume} {83}},\ \bibinfo
  {pages} {205101} (\bibinfo {year} {2011})}\BibitemShut {NoStop}%
\bibitem [{\citenamefont {Zhang}\ \emph {et~al.}(2021)\citenamefont {Zhang},
  \citenamefont {Xu}, \citenamefont {Carnahan}, \citenamefont {Sretenovic},
  \citenamefont {Suri}, \citenamefont {Xiao},\ and\ \citenamefont
  {Ke}}]{PhysRevLett.127.247202}%
  \BibitemOpen
  \bibfield  {author} {\bibinfo {author} {\bibfnamefont {H.}~\bibnamefont
  {Zhang}}, \bibinfo {author} {\bibfnamefont {C.}~\bibnamefont {Xu}}, \bibinfo
  {author} {\bibfnamefont {C.}~\bibnamefont {Carnahan}}, \bibinfo {author}
  {\bibfnamefont {M.}~\bibnamefont {Sretenovic}}, \bibinfo {author}
  {\bibfnamefont {N.}~\bibnamefont {Suri}}, \bibinfo {author} {\bibfnamefont
  {D.}~\bibnamefont {Xiao}},\ and\ \bibinfo {author} {\bibfnamefont
  {X.}~\bibnamefont {Ke}},\ }\bibfield  {title} {\bibinfo {title} {{Anomalous
  Thermal Hall Effect in an Insulating van der Waals Magnet}},\ }\href
  {https://doi.org/10.1103/PhysRevLett.127.247202} {\bibfield  {journal}
  {\bibinfo  {journal} {Phys. Rev. Lett.}\ }\textbf {\bibinfo {volume} {127}},\
  \bibinfo {pages} {247202} (\bibinfo {year} {2021})}\BibitemShut {NoStop}%
\bibitem [{\citenamefont {Li}\ \emph {et~al.}(2016)\citenamefont {Li},
  \citenamefont {Li}, \citenamefont {Kim}, \citenamefont {Balents},
  \citenamefont {Yu},\ and\ \citenamefont {Chen}}]{Li2016}%
  \BibitemOpen
  \bibfield  {author} {\bibinfo {author} {\bibfnamefont {F.-Y.}\ \bibnamefont
  {Li}}, \bibinfo {author} {\bibfnamefont {Y.-D.}\ \bibnamefont {Li}}, \bibinfo
  {author} {\bibfnamefont {Y.~B.}\ \bibnamefont {Kim}}, \bibinfo {author}
  {\bibfnamefont {L.}~\bibnamefont {Balents}}, \bibinfo {author} {\bibfnamefont
  {Y.}~\bibnamefont {Yu}},\ and\ \bibinfo {author} {\bibfnamefont
  {G.}~\bibnamefont {Chen}},\ }\bibfield  {title} {\bibinfo {title} {{Weyl
  magnons in breathing pyrochlore antiferromagnets}},\ }\href
  {https://doi.org/10.1038/ncomms12691} {\bibfield  {journal} {\bibinfo
  {journal} {Nature Communications}\ }\textbf {\bibinfo {volume} {7}},\
  \bibinfo {pages} {12691} (\bibinfo {year} {2016})}\BibitemShut {NoStop}%
\bibitem [{\citenamefont {Jian}\ and\ \citenamefont
  {Nie}(2018)}]{PhysRevB.97.115162}%
  \BibitemOpen
  \bibfield  {author} {\bibinfo {author} {\bibfnamefont {S.-K.}\ \bibnamefont
  {Jian}}\ and\ \bibinfo {author} {\bibfnamefont {W.}~\bibnamefont {Nie}},\
  }\bibfield  {title} {\bibinfo {title} {{Weyl magnons in pyrochlore
  antiferromagnets with an all-in-all-out order}},\ }\href
  {https://doi.org/10.1103/PhysRevB.97.115162} {\bibfield  {journal} {\bibinfo
  {journal} {Phys. Rev. B}\ }\textbf {\bibinfo {volume} {97}},\ \bibinfo
  {pages} {115162} (\bibinfo {year} {2018})}\BibitemShut {NoStop}%
\bibitem [{\citenamefont {Li}\ and\ \citenamefont
  {Nevidomskyy}(2021)}]{PhysRevB.104.104419}%
  \BibitemOpen
  \bibfield  {author} {\bibinfo {author} {\bibfnamefont {S.}~\bibnamefont
  {Li}}\ and\ \bibinfo {author} {\bibfnamefont {A.~H.}\ \bibnamefont
  {Nevidomskyy}},\ }\bibfield  {title} {\bibinfo {title} {{Topological Weyl
  magnons and thermal Hall effect in layered honeycomb ferromagnets}},\ }\href
  {https://doi.org/10.1103/PhysRevB.104.104419} {\bibfield  {journal} {\bibinfo
   {journal} {Phys. Rev. B}\ }\textbf {\bibinfo {volume} {104}},\ \bibinfo
  {pages} {104419} (\bibinfo {year} {2021})}\BibitemShut {NoStop}%
\bibitem [{\citenamefont {Chen}\ \emph {et~al.}(2018)\citenamefont {Chen},
  \citenamefont {Chung}, \citenamefont {Gao}, \citenamefont {Chen},
  \citenamefont {Stone}, \citenamefont {Kolesnikov}, \citenamefont {Huang},\
  and\ \citenamefont {Dai}}]{PhysRevX.8.041028}%
  \BibitemOpen
  \bibfield  {author} {\bibinfo {author} {\bibfnamefont {L.}~\bibnamefont
  {Chen}}, \bibinfo {author} {\bibfnamefont {J.-H.}\ \bibnamefont {Chung}},
  \bibinfo {author} {\bibfnamefont {B.}~\bibnamefont {Gao}}, \bibinfo {author}
  {\bibfnamefont {T.}~\bibnamefont {Chen}}, \bibinfo {author} {\bibfnamefont
  {M.~B.}\ \bibnamefont {Stone}}, \bibinfo {author} {\bibfnamefont {A.~I.}\
  \bibnamefont {Kolesnikov}}, \bibinfo {author} {\bibfnamefont
  {Q.}~\bibnamefont {Huang}},\ and\ \bibinfo {author} {\bibfnamefont
  {P.}~\bibnamefont {Dai}},\ }\bibfield  {title} {\bibinfo {title}
  {{Topological Spin Excitations in Honeycomb Ferromagnet
  ${\mathrm{CrI}}_{3}$}},\ }\href {https://doi.org/10.1103/PhysRevX.8.041028}
  {\bibfield  {journal} {\bibinfo  {journal} {Phys. Rev. X}\ }\textbf {\bibinfo
  {volume} {8}},\ \bibinfo {pages} {041028} (\bibinfo {year}
  {2018})}\BibitemShut {NoStop}%
\bibitem [{\citenamefont {Zhang}\ and\ \citenamefont
  {Yao}(2023)}]{PhysRevB.107.024408}%
  \BibitemOpen
  \bibfield  {author} {\bibinfo {author} {\bibfnamefont {M.-H.}\ \bibnamefont
  {Zhang}}\ and\ \bibinfo {author} {\bibfnamefont {D.-X.}\ \bibnamefont
  {Yao}},\ }\bibfield  {title} {\bibinfo {title} {{Topological magnons on the
  triangular kagome lattice}},\ }\href
  {https://doi.org/10.1103/PhysRevB.107.024408} {\bibfield  {journal} {\bibinfo
   {journal} {Phys. Rev. B}\ }\textbf {\bibinfo {volume} {107}},\ \bibinfo
  {pages} {024408} (\bibinfo {year} {2023})}\BibitemShut {NoStop}%
\bibitem [{\citenamefont {Liu}\ \emph {et~al.}(2023)\citenamefont {Liu},
  \citenamefont {Wang},\ and\ \citenamefont {Shen}}]{PhysRevB.107.174404}%
  \BibitemOpen
  \bibfield  {author} {\bibinfo {author} {\bibfnamefont {J.}~\bibnamefont
  {Liu}}, \bibinfo {author} {\bibfnamefont {L.}~\bibnamefont {Wang}},\ and\
  \bibinfo {author} {\bibfnamefont {K.}~\bibnamefont {Shen}},\ }\bibfield
  {title} {\bibinfo {title} {{Tunable topological magnon phases in layered
  ferrimagnets}},\ }\href {https://doi.org/10.1103/PhysRevB.107.174404}
  {\bibfield  {journal} {\bibinfo  {journal} {Phys. Rev. B}\ }\textbf {\bibinfo
  {volume} {107}},\ \bibinfo {pages} {174404} (\bibinfo {year}
  {2023})}\BibitemShut {NoStop}%
\bibitem [{\citenamefont {McClarty}(2022)}]{annurev}%
  \BibitemOpen
  \bibfield  {author} {\bibinfo {author} {\bibfnamefont {P.~A.}\ \bibnamefont
  {McClarty}},\ }\bibfield  {title} {\bibinfo {title} {{Topological Magnons: A
  Review}},\ }\href {https://doi.org/10.1146/annurev-conmatphys-031620-104715}
  {\bibfield  {journal} {\bibinfo  {journal} {Annual Review of Condensed Matter
  Physics}\ }\textbf {\bibinfo {volume} {13}},\ \bibinfo {pages} {171}
  (\bibinfo {year} {2022})}\BibitemShut {NoStop}%
\bibitem [{\citenamefont {Joshi}(2018)}]{PhysRevB.98.060405}%
  \BibitemOpen
  \bibfield  {author} {\bibinfo {author} {\bibfnamefont {D.~G.}\ \bibnamefont
  {Joshi}},\ }\bibfield  {title} {\bibinfo {title} {{Topological excitations in
  the ferromagnetic Kitaev-Heisenberg model}},\ }\href
  {https://doi.org/10.1103/PhysRevB.98.060405} {\bibfield  {journal} {\bibinfo
  {journal} {Phys. Rev. B}\ }\textbf {\bibinfo {volume} {98}},\ \bibinfo
  {pages} {060405} (\bibinfo {year} {2018})}\BibitemShut {NoStop}%
\bibitem [{\citenamefont {Xu}\ \emph {et~al.}(2020)\citenamefont {Xu},
  \citenamefont {Flynn}, \citenamefont {Alase}, \citenamefont {Cobanera},
  \citenamefont {Viola},\ and\ \citenamefont {Ortiz}}]{PhysRevB.102.125127}%
  \BibitemOpen
  \bibfield  {author} {\bibinfo {author} {\bibfnamefont {Q.-R.}\ \bibnamefont
  {Xu}}, \bibinfo {author} {\bibfnamefont {V.~P.}\ \bibnamefont {Flynn}},
  \bibinfo {author} {\bibfnamefont {A.}~\bibnamefont {Alase}}, \bibinfo
  {author} {\bibfnamefont {E.}~\bibnamefont {Cobanera}}, \bibinfo {author}
  {\bibfnamefont {L.}~\bibnamefont {Viola}},\ and\ \bibinfo {author}
  {\bibfnamefont {G.}~\bibnamefont {Ortiz}},\ }\bibfield  {title} {\bibinfo
  {title} {{Squaring the fermion: The threefold way and the fate of zero
  modes}},\ }\href {https://doi.org/10.1103/PhysRevB.102.125127} {\bibfield
  {journal} {\bibinfo  {journal} {Phys. Rev. B}\ }\textbf {\bibinfo {volume}
  {102}},\ \bibinfo {pages} {125127} (\bibinfo {year} {2020})}\BibitemShut
  {NoStop}%
\bibitem [{\citenamefont {Rosenberg}\ and\ \citenamefont
  {Manousakis}(2022)}]{PhysRevB.106.054511}%
  \BibitemOpen
  \bibfield  {author} {\bibinfo {author} {\bibfnamefont {P.}~\bibnamefont
  {Rosenberg}}\ and\ \bibinfo {author} {\bibfnamefont {E.}~\bibnamefont
  {Manousakis}},\ }\bibfield  {title} {\bibinfo {title} {{Topological
  superconductivity in a two-dimensional Weyl SSH model}},\ }\href
  {https://doi.org/10.1103/PhysRevB.106.054511} {\bibfield  {journal} {\bibinfo
   {journal} {Phys. Rev. B}\ }\textbf {\bibinfo {volume} {106}},\ \bibinfo
  {pages} {054511} (\bibinfo {year} {2022})}\BibitemShut {NoStop}%
\bibitem [{\citenamefont {Rosenberg}\ and\ \citenamefont
  {Manousakis}(2021)}]{PhysRevB.104.134511}%
  \BibitemOpen
  \bibfield  {author} {\bibinfo {author} {\bibfnamefont {P.}~\bibnamefont
  {Rosenberg}}\ and\ \bibinfo {author} {\bibfnamefont {E.}~\bibnamefont
  {Manousakis}},\ }\bibfield  {title} {\bibinfo {title} {{Weyl nodal-ring
  semimetallic behavior and topological superconductivity in crystalline forms
  of Su-Schrieffer-Heeger chains}},\ }\href
  {https://doi.org/10.1103/PhysRevB.104.134511} {\bibfield  {journal} {\bibinfo
   {journal} {Phys. Rev. B}\ }\textbf {\bibinfo {volume} {104}},\ \bibinfo
  {pages} {134511} (\bibinfo {year} {2021})}\BibitemShut {NoStop}%
\bibitem [{\citenamefont {Colpa}(1978)}]{COLPA1978327}%
  \BibitemOpen
  \bibfield  {author} {\bibinfo {author} {\bibfnamefont {J.}~\bibnamefont
  {Colpa}},\ }\bibfield  {title} {\bibinfo {title} {{Diagonalization of the
  quadratic boson hamiltonian}},\ }\href
  {https://doi.org/https://doi.org/10.1016/0378-4371(78)90160-7} {\bibfield
  {journal} {\bibinfo  {journal} {Physica A: Statistical Mechanics and its
  Applications}\ }\textbf {\bibinfo {volume} {93}},\ \bibinfo {pages} {327}
  (\bibinfo {year} {1978})}\BibitemShut {NoStop}%
\bibitem [{\citenamefont {Simon}\ \emph {et~al.}(2011)\citenamefont {Simon},
  \citenamefont {Bakr}, \citenamefont {Ma}, \citenamefont {Tai}, \citenamefont
  {Preiss},\ and\ \citenamefont {Greiner}}]{Simon2011}%
  \BibitemOpen
  \bibfield  {author} {\bibinfo {author} {\bibfnamefont {J.}~\bibnamefont
  {Simon}}, \bibinfo {author} {\bibfnamefont {W.~S.}\ \bibnamefont {Bakr}},
  \bibinfo {author} {\bibfnamefont {R.}~\bibnamefont {Ma}}, \bibinfo {author}
  {\bibfnamefont {M.~E.}\ \bibnamefont {Tai}}, \bibinfo {author} {\bibfnamefont
  {P.~M.}\ \bibnamefont {Preiss}},\ and\ \bibinfo {author} {\bibfnamefont
  {M.}~\bibnamefont {Greiner}},\ }\bibfield  {title} {\bibinfo {title}
  {{Quantum simulation of antiferromagnetic spin chains in an optical
  lattice}},\ }\href {https://doi.org/10.1038/nature09994} {\bibfield
  {journal} {\bibinfo  {journal} {Nature}\ }\textbf {\bibinfo {volume} {472}},\
  \bibinfo {pages} {307} (\bibinfo {year} {2011})}\BibitemShut {NoStop}%
\bibitem [{\citenamefont {Brown}\ \emph {et~al.}(2017)\citenamefont {Brown},
  \citenamefont {Mitra}, \citenamefont {Guardado-Sanchez}, \citenamefont
  {Schauß}, \citenamefont {Kondov}, \citenamefont {Khatami}, \citenamefont
  {Paiva}, \citenamefont {Trivedi}, \citenamefont {Huse},\ and\ \citenamefont
  {Bakr}}]{doi:10.1126/science.aam7838}%
  \BibitemOpen
  \bibfield  {author} {\bibinfo {author} {\bibfnamefont {P.~T.}\ \bibnamefont
  {Brown}}, \bibinfo {author} {\bibfnamefont {D.}~\bibnamefont {Mitra}},
  \bibinfo {author} {\bibfnamefont {E.}~\bibnamefont {Guardado-Sanchez}},
  \bibinfo {author} {\bibfnamefont {P.}~\bibnamefont {Schauß}}, \bibinfo
  {author} {\bibfnamefont {S.~S.}\ \bibnamefont {Kondov}}, \bibinfo {author}
  {\bibfnamefont {E.}~\bibnamefont {Khatami}}, \bibinfo {author} {\bibfnamefont
  {T.}~\bibnamefont {Paiva}}, \bibinfo {author} {\bibfnamefont
  {N.}~\bibnamefont {Trivedi}}, \bibinfo {author} {\bibfnamefont {D.~A.}\
  \bibnamefont {Huse}},\ and\ \bibinfo {author} {\bibfnamefont {W.~S.}\
  \bibnamefont {Bakr}},\ }\bibfield  {title} {\bibinfo {title} {{Spin-imbalance
  in a 2D Fermi-Hubbard system}},\ }\href
  {https://doi.org/10.1126/science.aam7838} {\bibfield  {journal} {\bibinfo
  {journal} {Science}\ }\textbf {\bibinfo {volume} {357}},\ \bibinfo {pages}
  {1385} (\bibinfo {year} {2017})}\BibitemShut {NoStop}%
\end{thebibliography}
\end{document}